\documentstyle[aps,preprint]{revtex}

\begin{document}
\draft
\title{Could the Gravitons Have Rest Masses?}
\author{Liao Liu\thanks{%
e-mail: liuliao1928@yahoo.com.cn}, Zheng Zhao}
\address{Department of Physics, Institute of Theoretical Physics, Beijing Normal\\
University, Beijing 100875, China}
\maketitle

\begin{abstract}
If the cosmological constant $\Lambda $ can not be neglected, we will show
in this short report, that graviton should have a rest mass $m_g=\sqrt{%
2\Lambda }=\sqrt{2\Lambda }\hbar c^{-1}$ different from zero.

{\bf Keywords:} graviton; cosmology; rest mass

{\bf PACS numbers:} 97.60.Lf
\end{abstract}

As was shown by S. Perlmutter et al in 1999$^{[1]}$, that a best fit flat
cosmology to the observed Hubble constant $H_0$ and cosmological constant $%
\Lambda $ from the red-shift of Ia-type supernova can be obtained, that is

\begin{eqnarray*}
\Omega _m+\Omega _\Lambda &=&1, \\
(\Omega _m,\Omega _\Lambda ) &=&(0.28,0.72),
\end{eqnarray*}
wherein $\Omega _{m,\Lambda }\equiv \frac{\rho _{m,\Lambda }}{\rho _c},$ $%
\rho _c\equiv \frac{3H_0^2}{8\pi G}.$

The above result suggests strongly that the so called ''cosmological vacuum
energy density $\frac \Lambda {8\pi G}=\rho _\Lambda $'' is about 2.6 times
that of the matter field in our universe! Henceforth we have no right to
ignore the cosmological term in Einstein's gravitational field equation,
though Einstein said it is ''the biggest blunder of my life''. Now, let us
consider the linear approximation of weak gravitational field, in this
approximation we can write the gravitational field potential $g_{\mu \nu }$
as

\begin{equation}
g_{\mu \nu }=G_{\mu \nu }+h_{\mu \nu },  \label{eq1}
\end{equation}
where $G_{\mu \nu }$ is the back-ground Minkovsky flat metric, $h_{\mu \nu }$
is the derivation from $G_{\mu \nu }$, such that $h_{\mu \nu }$ and their
derivatives are all small whose squares may be neglected.

It is straightforward to get the linear approximation of Einstein's
gravitational field equation with cosmological term as follows$^{[2]}$

\begin{equation}
\Box h_{\mu \nu }-2\Lambda (G_{\mu \nu }+h_{\mu \nu })=-16\pi G(T_{\mu \nu
}^{(m)}-\frac 12T^{(m)}g_{\mu \nu })\equiv -16\pi G\Gamma _{\mu \nu }^{(m)},
\label{eq2}
\end{equation}
or 
\begin{equation}
\Box h_{\mu \nu }-2\Lambda h_{\mu \nu }=-16\pi G(\Gamma _{\mu \nu
}^{(m)}+T_{\mu \nu }^{(\Lambda )})\equiv -16\pi G(T_{\mu \nu }),  \label{eq3}
\end{equation}
where $\Gamma _{\mu \nu }^{(m)}$ belongs to the matter fields, $T_{\mu \nu
}^{(\Lambda )}\equiv -\frac \Lambda {8\pi G}G_{\mu \nu }$ belongs to the
cosmological constant. Eq.(\ref{eq3}) is really a Lorentzian covariant
tensor field equation with source term $T_{\mu \nu }$ and mass term $%
(2\Lambda )h_{\mu \nu }$ for $h_{\mu \nu }$, which implies that the
gravitational field in linear approximation can be look upon as a spin 2
particle with rest mass $\sqrt{2\Lambda }\hbar c^{-1}$. Especially in static
case Eq.(\ref{eq3}) becomes$^{[2]}$

\begin{equation}
\triangle \varphi -2\Lambda \varphi =4\pi G\stackrel{\symbol{94}}{\rho },
\label{eq4}
\end{equation}
where $\stackrel{\symbol{94}}{\rho }\equiv \rho _m+\frac \Lambda {4\pi G}$, $%
g_{00}=G_{00}+h_{00}=1+2\varphi $, $\varphi $ is the gravitational potential.

We remark that Eq.(\ref{eq4}) is just the famous Seeliger gravitational
field equation found by C. Neumann and H. V. Seeliger$^{[3-5]}$ long ago.
Now let us try to decompose Eq.(\ref{eq4}) and find out their associated
solutions as follows

\begin{equation}
\triangle \varphi _m(r)-(2\Lambda )\varphi _m(r)=4\pi G\rho _m,  \label{eq5}
\end{equation}
\begin{equation}
\triangle \varphi _\Lambda -(2\Lambda )\varphi _\Lambda =4\pi G\left( \frac %
\Lambda {4\pi G}\right) ,  \label{eq6}
\end{equation}
\[
\varphi =\varphi _m+\varphi _\Lambda .
\]
A special solution of Eq.(\ref{eq6}) is

\begin{equation}
\varphi _\Lambda =-\frac 12=\text{cons}\tan \text{t,}  \label{eq7}
\end{equation}
but the solution of Eq.(\ref{eq5}) is a Yukawa-type potential 
\begin{equation}
\varphi _m=-\frac{Gm}re^{-\sqrt{2\Lambda }\cdot r},\rho _m(r)=m\delta
(r-r_0),  \label{eq8}
\end{equation}
the combined potential 
\[
\varphi =\varphi _m+\varphi _\Lambda 
\]
is certainly a solution of the combined field equation(\ref{eq4}). Since the
constant potential $\varphi _\Lambda $ has no contribution to the field
strength, so only the Yukawa potential $\varphi _m$ has physical
significance. Evidently now the Newton universal gravitational force law
should be changed to the longly forgottened Newmann-Seeliger's law 
\begin{equation}
f=-G\frac{m_1m_2}{r^2}e^{-\sqrt{2\Lambda }\cdot r}.  \label{eq9}
\end{equation}
So the very surprised thing is that the re-confirmation of cosmological
constant $\Lambda \neq 0$ compels us to substitute the Newton's celebrated
law by the Newmann-Seeliger law!

It seems, moreover our report may make clear that there is no such things as
''dark energy'' or ''gravitational vacuum energy'' in our universe, rather
than, the cosmological term just tell us that graviton has a rest mass $%
\sqrt{2\Lambda }\hbar c^{-1}$ different from zero! As pointed by J. N. Islam$%
^{[6]}$ that $|\Lambda |$ has an upper limit of 10$^{-42}$cm$^{-2}$
nowadays, then the rest mass of graviton should also have an upper limit 10$%
^{-58}$g, it is too small to be detected. However, it is believed that our
very early universe may have a very large cosmological constant, say $%
\Lambda \simeq 10^{34}$cm$^{-2}$ $^{[7]}$, then the rest mass of graviton is
10$^{-20}$g. This implies that graviton created in very early universe or
from very distant astronomical incidences may have a measurable rest mass.

We are grateful to the discussion with Prof. C. G. Huang, though he didn't
agree with our point view. Many helps from Dr. W. B. Liu are acknowledged.

This work is supported by the National Natural Science Foundation of China
under Grant No. 10373003.

\section*{References}

\begin{enumerate}
\item  \label{r1}S Perlmutter, et al APJ 517(1999) 565

\item  C M$\phi $ller The theory of relativity(1955) 118, 119

\item  C Newmann, Leipziger Ahhandlungen(1874)

\item  H V Seeliger Vietel-Jahresschrift des Astronomischen Gesselschaft
41(1906 )234

\item  M V Laue Die Relativit$\stackrel{..}{a}$ts theories(1956) Band II,
Sect 11

\item  J N Islam Phys Lett 97A(1983) 239

\item  L Liu, S Y Pei Chin Phys Lett\label{r14} 5(2003)774
\end{enumerate}

\end{document}